\documentclass[prl,twocolumn,letterpaper,superscriptaddress,amsmath,amssymb,showpacs,longbibliography]{revtex4-1}

 \pdfoutput=1

\usepackage{graphicx}
\usepackage{color}
\usepackage{epsfig}
\usepackage{subfigure}
\usepackage{natbib}
\usepackage{array}
\usepackage{amsmath}
\usepackage{ulem}

\usepackage{tikz}

\newcommand*{\circled}[2][]{\tikz[baseline=(C.base)]{
    \node[inner sep=0pt] (C) {\vphantom{1g}#2};
    \node[draw, circle, inner sep=2pt, yshift=1pt] 
        at (C.center) {\vphantom{1g}};}}

\makeatletter

\begin{document}
\newcolumntype{M}{>{$}c<{$}}


\title{Site-dependent properties of quantum emitters in nanostructured silicon carbide} 

\author{Tamanna Joshi}

\affiliation{Department of Physics and Astronomy, Howard University, Washington, D.C. 20059, USA}

\author{Pratibha Dev}

\affiliation{Department of Physics and Astronomy, Howard University, Washington, D.C. 20059, USA}
\date{\today}

\begin{abstract}

Deep defects in silicon carbide (SiC) possess atom-like electronic, spin and optical properties, making them ideal for quantum-computing and -sensing applications. In these applications, deep defects are often placed within fabricated nanostructures that modify defect properties due to surface and quantum confinement effects. Thus far, theoretical studies exploring deep defects in SiC have ignored these effects. Using density functional theory, this work demonstrates site-dependence of properties of bright, negatively-charged silicon monovacancies within a SiC nanowire. It is shown that the optical properties of defects depend strongly on the hybridization of the defect states with the surface states and on the structural changes allowed by proximity to the surfaces. Additionally, the analysis of the first principles results indicates that the charge-state conversion and/or migration to thermodynamically-favorable undercoordinated surface sites can deteriorate deep-defect properties. These results illustrate the importance of considering how finite-size effects tune defect properties, and of creating mitigating protocols to ensure a defect's charge-state stability within nanostructured hosts.



	\end{abstract}
\maketitle

\section{Introduction}
Silicon carbide (SiC) hosts a number of bright deep defects that are promising for use as quantum bits (qubits) in quantum information~\cite{weber2011defects,Carter2015,economou2016spin,nagy2018quantumdichroic,lukinRadulaski20204h,lukin2020siliconvacancy}, and quantum-enhanced sensing applications~\cite{niethammer2016vectorMagnetometry}. The quantum emitters in SiC have addressable spins~\cite{falk2013polytype, Seo2016,nagy2019highfidelity}, long spin coherence times~\cite{Christle2015,Seo2016} and near-telecom range quantum emissions~\cite{Christle2015, Fuchs2015}. But above all, what gives SiC an edge over other wide bandgap hosts such as diamond, is that it is an industrially-mature material, allowing for nanofabrication and scalability. It also exists in a number of low-energy cubic (3C-SiC) and hexagonal (2H-, 4H-, and 6H-SiC) polytypes, providing a practical and large playing field for implementing different technologies.

Notwithstanding the progress made in the field, the exploration of SiC as a host to quantum emitters is a relatively new endeavour~\cite{weber2010quantum}, and there still are challenges in integrating quantum emitters in SiC into photonics and in optimizing the generation of the deep defects within SiC~\cite{Castelletto_Review_2020}. In addition, very few experiments~\cite{Polking_deLeon_SiC_surface2018,nanostr_VSi_expt_2020} and no theoretical studies (to the best of our knowledge), which are usually performed in bulk SiC, have probed surface and/or finite size-effects, even though shallower defects are desirable for sensing applications.  These effects also become important when the quantum emitters are placed in optical nanophotonic cavities~\cite{gadalla2021enhanced, majety2021quantum, crookEHu2020purcell, zhangRadulaskiMarko2018strongly, bracherEHu2017selective}, nanoscale solid immersion lenses~\cite{widmann2015coherentRT}, and arrays of nanopillars~\cite{lukinRadulaski20204h,nagy2018quantumdichroic,radulaski2017scalable}. In such nanofabricated devices, inevitably there will be defects close to surfaces or sidewalls. This can change, or even deteriorate a defect's optical properties, and make the bright (photoluminescent) charge state unstable as different finite-size effects come into play within a nanostructure~\cite{lukinRadulaski20204h,YuanDeLeon2020Charge,sangtawesinDeLeon2019origins}. 
In the present work, we have investigated the effects of nanostructuring on structural, electronic, spin and optical properties of defects using density functional theory (DFT).  In particular, we show how quantum confinement and surface effects influence the properties of the bright, negatively charged silicon-vacancy (V$_\text{Si}^{-1}$) by placing these defects in a 2H-SiC nanowire (NW).  We have chosen V$_\text{Si}^{-1}$ as our proof-of-principle defect because it is a particularly noteworthy qubit candidate; it shows an optically-accessible high spin state (S=3/2)~\cite{Carter2015,soykal2016silicon}, a high emission into the zero phonon line (ZPL)~\cite{nagy2018quantumdichroic}, spectral stability~\cite{Udvarhelyi2019}, and long, room temperature spin coherence~\cite{widmann2015coherentRT, Carter2015, nagy2019highfidelity}. 
Our results demonstrate that the deterioration of the optical signal from the bright monovacancies close to the surfaces can be attributed to conversion to a dark charge-state and/or defect migration to the thermodynamically-favorable undercoordinated surface sites, which act as a sink to the defects. These results illustrate the importance of: (i) considering how finite-size effects can change and hence, be leveraged to tune defects' properties by deterministic placement of the defects at different distance from the surfaces via use of different implantation techniques~\cite{Wang_Depth_10nm_2017,Carter_VSi_2021}, and (ii) creating mitigating protocols to ensure a defect's charge-state stability within nanostructured hosts via different surface passivation schemes. 


\section{Computational Details}

\begin{figure*}[ht]
	\centering
	\includegraphics[width=0.7\linewidth]{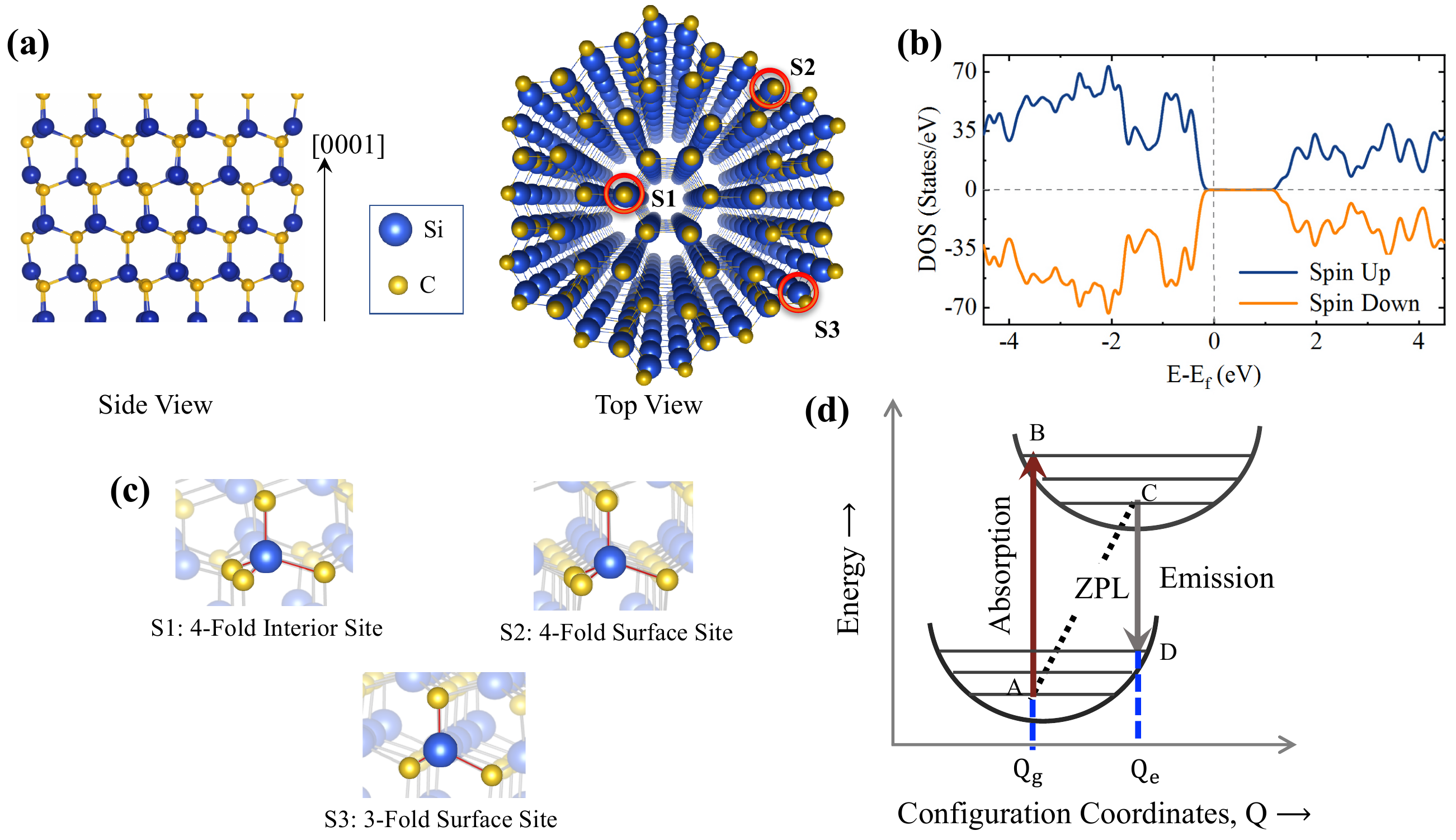}
	\caption{2H-SiC Nanowire: (a) Side- and top-view of pristine/defect-free NW, periodic along the [0001]-direction. (b) Spin-resolved density of states (DOS) plot for the pristine NW, showing that it is non-magnetic. (c) The three distinct silicon sites [circled in (a)] at which the defects were created are: the tetrahedrally bonded interior site (S1), the 4-fold coordinated surface site (S2), and the under-coordinated site on the surface, bonded to only three carbon atoms (S3). (d) The configuration coordinate diagram, giving total energy as a function of ground (bottom) and excited state (top) configurations within the Franck Condon picture. It is used to emulate the photoexcitation process, which is calculated using the constrained-occupation DFT (CDFT) method.}
	\label{fig:SiC-NW-Structure}
\end{figure*}

  All of the first principle density functional theory (DFT) based calculations used the generalized gradient approximation (GGA)~\cite{GGA} of Perdew-Burke-Ernzerhof ~\cite{PBE} implemented within the Quantum Espresso package~\cite{QE-2009, QE-2017}. As we used ultrasoft pseudopotentials, we found that cutoffs of 40\,Ry and 350\,Ry were sufficient for expanding the wavefunctions and charge-densities.  All structures were fully relaxed till the forces on atoms were below 10$^\text{-3}$\,Ry/a.u. 
 
 
In this work, we chose the 2H-SiC polytype (over the 4H- and 6H-SiC polytypes) as its allowed us to create a non-magnetic nanostructure. Nevertheless, our qualitative results are general and should apply to other choices of SiC polytypes as well, even if finer quantitative details may differ due to the different stacking sequences of the tetrahedrally-bonded silicon- and carbon-atoms along the c-axis. The 2H-SiC NW, shown in Fig.~\ref{fig:SiC-NW-Structure}(a), is periodic in the [0001]-direction and consists of 216 atoms.  Periodic images of the NW in the lateral directions were separated by a vacuum of about 12\AA{}. Brillouin zone sampling is carried out by using a $\Gamma$-centered k-point grid of $1 \times 1 \times 6$ according to the Monkhorst-Pack scheme~\cite{Monkhorst}. The as-created pristine NW is stoichiometric and non-magnetic. This can be seen in Fig.~\ref{fig:SiC-NW-Structure}(b), which shows the spin-resolved density of states (DOS) with completely symmetric contributions from the two spin channels. The absence of magnetic defects in the pristine NW ensured that once silicon vacancies were introduced, there were no additional interactions between the spins of silicon vacancies and those of magnetic surface defects. Although in experiments, there can be other magnetic defects in the interior and on the surface of such a nanostructure, our choice of a non-magnetic pristine nanowire allowed us to concentrate on the role of finite-size effects by themselves in changing the properties of deep defects, once the latter were introduced.

In the equilibrium structure of the pristine NW, there are three distinct NW sites at which we created a silicon-vacancy [see Figs~\ref{fig:SiC-NW-Structure}(a) and (c)]: (i) 4-fold coordinated Si-atoms in the interior of the NW ($\text{S1}$), (ii) 4-fold coordinated Si-atoms on the NW surface ($\text{S2}$), and (iii) 3-fold site on the surface ($\text{S3}$). Since the defects can undergo charge conversion from the bright (i.e. photoluminescent) negatively charged-state to the dark neutral state upon photoexcitation, we studied the properties of V$_\text{Si}$ in both of these charged states.  The formation energy, $\Delta E^{form}$, of the defect $X$ in charged state $q$ is defined as:

\begin{equation}
\begin{aligned}
\Delta E^{form}[X^q] = {} & E_t[X^q]-E_t[\text{prist.}]-\sum_i n_i\mu_i \\
&+q\left( E_v+E_F+\Delta V\right)
\end{aligned}
\end{equation}

\noindent where $E_t[X^q]$ is the total energy of the defective system, $E_t[\text{prist.}]$ is the total energy of the pristine/defect-free system, $q$ is the charged state of the defect, $E_v$ is the valance band maximum, $E_F$ is the Fermi energy referenced to the valance band maximum, $\Delta V$ is the potential alignment term, $\mu_i$ is the chemical potential of the $i^{th}$ atom, and the integer, $n_i$, represents the number of atoms added (a positive integer) or removed (negative integer).

\begin{figure*}[!ht]
	\centering
	\includegraphics[width=0.7\linewidth]{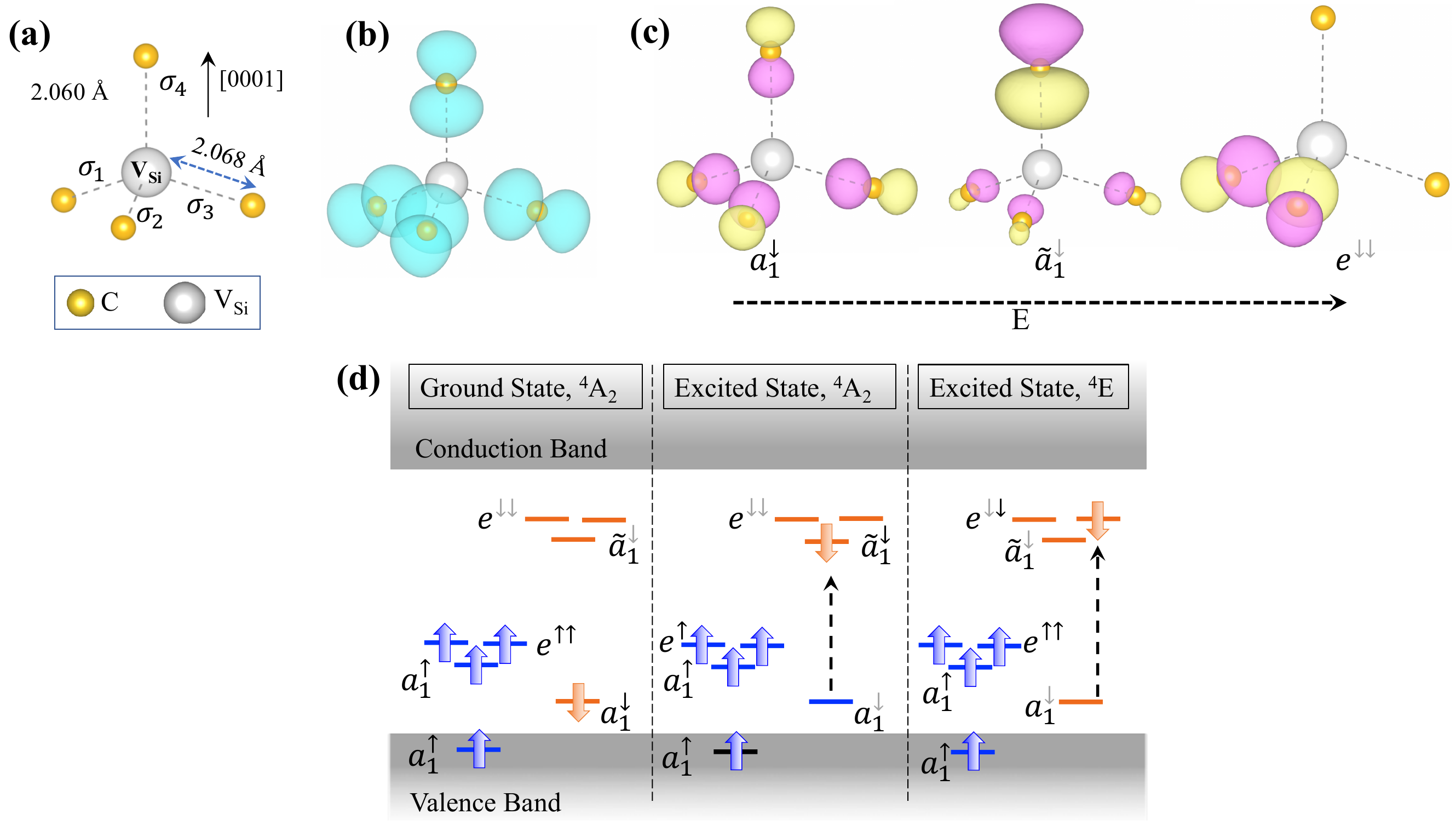}
	\caption{Charged silicon monovacancy in bulk 2H-SiC: (a) $C_{3v}$-point group symmetry of $\text{V}_{\text{Si}}^{-1}$. Only the nearest-neighbouring carbon atoms are shown for clarity  (b) Spin density [difference in charge densities within the two spin channels] isosurface plot (in blue), showing the spatially localized nature of the defect-induced spins.  (c) Charge density plots for the optically-active defect states (minority-spin) that are involved in the photoexcitation process. From left to right, lowest-energy filled singlet state ($a_{1}$-symmetry), unoccupied singlet, $\tilde{a}_{1}$, and one of the doubly-degenerate unoccupied $e$-states.  Pink (yellow) color corresponds to positive (negative) isovalues. The defect states are labelled by their respective group theoretical representations. Black (grey) arrows in the superscripts of the labels are used to emphasize that the states are filled (empty). (d) Schematic single particle level diagram (not to scale), showing the defect states introduced by $\text{V}_{\text{Si}}^{-1}$ in the ground-state and the two excited states.}
	\label{fig:Bulk_SiC}
\end{figure*}

Since 2H-SiC is a lesser-studied polytype as compared to other hexagonal polytypes, we also determined the properties of monovacancies in the bulk in order to highlight the effects of nanostructuring.  The bulk calculations were carried out on an $8\times8\times2$ supercell, consisting of 512-atoms. The size of the bulk supercell ensured that the distances of the defects from their periodic images in the lateral and vertical directions were the same/similar for the NW and the bulk supercell.  Even though the the bulk supercell is large, we used a $\Gamma$-centered $2\times 2 \times 6$ k-grid. In our previous works on negatively-charged silicon vacancies in a bulk 4H-SiC supercell of a similar size~\cite{Carter2015,soykal2016silicon,economou2016spin}, we found that a finite-size k-grid is necessary to accurately describe the defect properties, and that using only a $\Gamma$-point calculation may lead to erroneous results.

In order to study the effect of nanostructuring on the excited state properties, we calculated the zero phonon line (ZPL) for V$_\text{Si}^{-1}$ in the NW, and in the bulk using the constrained-occupation DFT (CDFT) method [see Fig.~\ref{fig:SiC-NW-Structure}(d)].  Within the  CDFT method, the spin-preserving photoluminescence process is emulated by constraining the occupation of the defect-states in the optically-active spin-channel.  The CDFT results are then mapped onto the Franck-Condon picture, giving optical transition energies~\cite{Gali_DeltaSCF,PDEV_hBN_2020}. 
In this picture, the vertical transitions from A$\rightarrow$B and C$\rightarrow$D in Fig.~\ref{fig:SiC-NW-Structure}(d) correspond to the photon absorption (excitation) and emission (deexcitation) processes, respectively. The difference between the total energies of the relaxed structures in the excited-state [point C] and  ground-state [point A] correspond to the ZPL energy. The Stokes shift is calculated by finding the difference in the vertical excitation (absorption) and the ZPL.

\section{Results}

\subsection{Negatively charged silicon vacancy in bulk}

 In the bulk 2H-SiC, a $\text{V}_{\text{Si}}^{-1}$ leaves a total of five electrons in four $sp^3$-hybridized dangling bonds on the surrounding carbon-atoms. The resulting equilibrium structure has a $C_{3v}$-point group symmetry as seen in Fig.~\ref{fig:Bulk_SiC}(a) [only nearest-neighbouring (NN) carbons are shown for clarity]. It should be pointed out that the defective structure has only weakly broken $T_d$ symmetry, with the distances from the defect site to the three equivalent carbon-atoms in the basal plane and to the carbon on the $C_3$-axis (along the crystalline [0001]-direction) differing by a mere $0.8\times10^{-2}$\,\AA{}. Our calculations show that in the bulk crystal, $\text{V}_\text{Si}^ {-1}$ is a spin-3/2 defect, consistent with the experimental and theoretical results obtained for other polytypes of SiC~\cite{Carter2015,soykal2016silicon,economou2016spin}. Figure~\ref{fig:Bulk_SiC}(b) is an isosurface plot of the spin density [$\Delta\rho=\rho^{\uparrow}-\rho^{\downarrow}$, $\rho$ = charge density] for $\text{V}_{\text{Si}}^{-1}$, showing that  the defect-induced spins are mostly localized on the four NN carbons. This is due to the highly localized nature of the $2s$- and $2p$-orbitals that form the dangling bonds on the carbons surrounding the defects. The large exchange interactions between the spatially-localized unpaired electrons in the defect states leads to the observed spin-polarized structures, which is also seen in deep defects involving the $2s$ and $2p$-derived defect states of other second row elements~\cite{Dev_PRL_DeepDefects_2008, Dev_PRB_DeepDefects_2010, Dev_PRB_NW_2010, PDEV_hBN_2020}. This means that there are more electrons in the majority-spin channel (spin-up) than the minority-spin channel (spin-down). In order to further understand the electronic, spin and optical properties of $\text{V}_{\text{Si}}^{-1}$, we formed single-electron molecular orbitals (MOs) for the defect states. The MOs are created from the symmetry-adapted linear combinations of three equivalent dangling sigma-bonds ($\sigma_{1}$, $\sigma_{2}$, and $\sigma_{3}$) from the carbon-atoms in the basal plane and the orbital, $\sigma_{4}$, belonging to the carbon atom on the $C_{3}$-axis. These symmetry-adapted MOs include two singlet states, $a_{1} = c_{1} (\sigma_{1}+\sigma_{2}+\sigma_{3})+c_{2} \sigma_{4}$ and $\tilde{a}_{1} = \tilde{c}_{1} ((\sigma_{1}+\sigma_{2}+\sigma_{3})+ \tilde{c}_{2} \sigma_{4}$, and the doubly-degenerate states, $e=\{e_X = c_3 (2\sigma_{3}-\sigma_{1}-\sigma_{2})$, $e_Y = c_4 (\sigma_{1}-\sigma_{2})$\}.
Here, we have labelled the defect states by their symmetry, and the coefficients [$c_{i}$/$\tilde{c}_{i}$] ensure that the MOs are normalized. The ``tilde" on  the molecular orbital, $\tilde{a}_{1}$ is placed to distinguish it from the other defect state also belonging to the $a_{1}$ representation. The energy ordering of these states can be obtained from the DFT calculations.  Figure~\ref{fig:Bulk_SiC}(c) plots the energy-ordered defect states (optically-active minority spin channel only) in the ground state. This ordering of the defect states implies that the ground state has the following electronic configuration: $a_{1}^{\uparrow\downarrow}\tilde{a}_{1}^{\uparrow}e^{\uparrow\uparrow}$.  As the structure has only a weakly-broken $T_d$-symmetry, the doubly degenerate $e$-states are only about 0.01\,eV above the $\tilde{a}_{1}$ states. This means that the spin-preserving optical excitations from the ground state to the two possible excited states will have very similar energies. Figure~\ref{fig:Bulk_SiC}(d) is a schematic diagram (not to scale), showing (from left to right) the electronic configurations in the ground state ($^{4}A_{2}$-symmetry), and the two possible excited states with $^{4}A_{2}$-symmetry and $^{4}E$-symmetry.  We have used capital letters for the group-theoretic representations of the many-body ground and excited states. The lowest-energy optical excitation is from the ground state, $^{4}A_{2}$ [$a_{1}^{\uparrow\downarrow}a_{1}^{\uparrow}e^{\uparrow\uparrow}$] to the excited $^{4}A_{2}$-state  [$a_{1}^{\uparrow}a_{1}^{\uparrow\downarrow}e^{\uparrow\uparrow}$]. For this excitation, the zero-phonon line (ZPL) energy is calculated to be 1.19\,eV.  Our results presented thus far for the ground and excited state properties of $\text{V}_{\text{Si}}^{-1}$ in the bulk 2H-SiC are consistent with those obtained for other polytypes of SiC~\cite{soykal2016silicon}. It is expected that the nanostructuring of the host crystal will modify the bulk properties. These modifications are critical to understanding experiments on nanostructured SiC. 


\begin{figure*}
	\centering
	\includegraphics[width=0.80\linewidth]{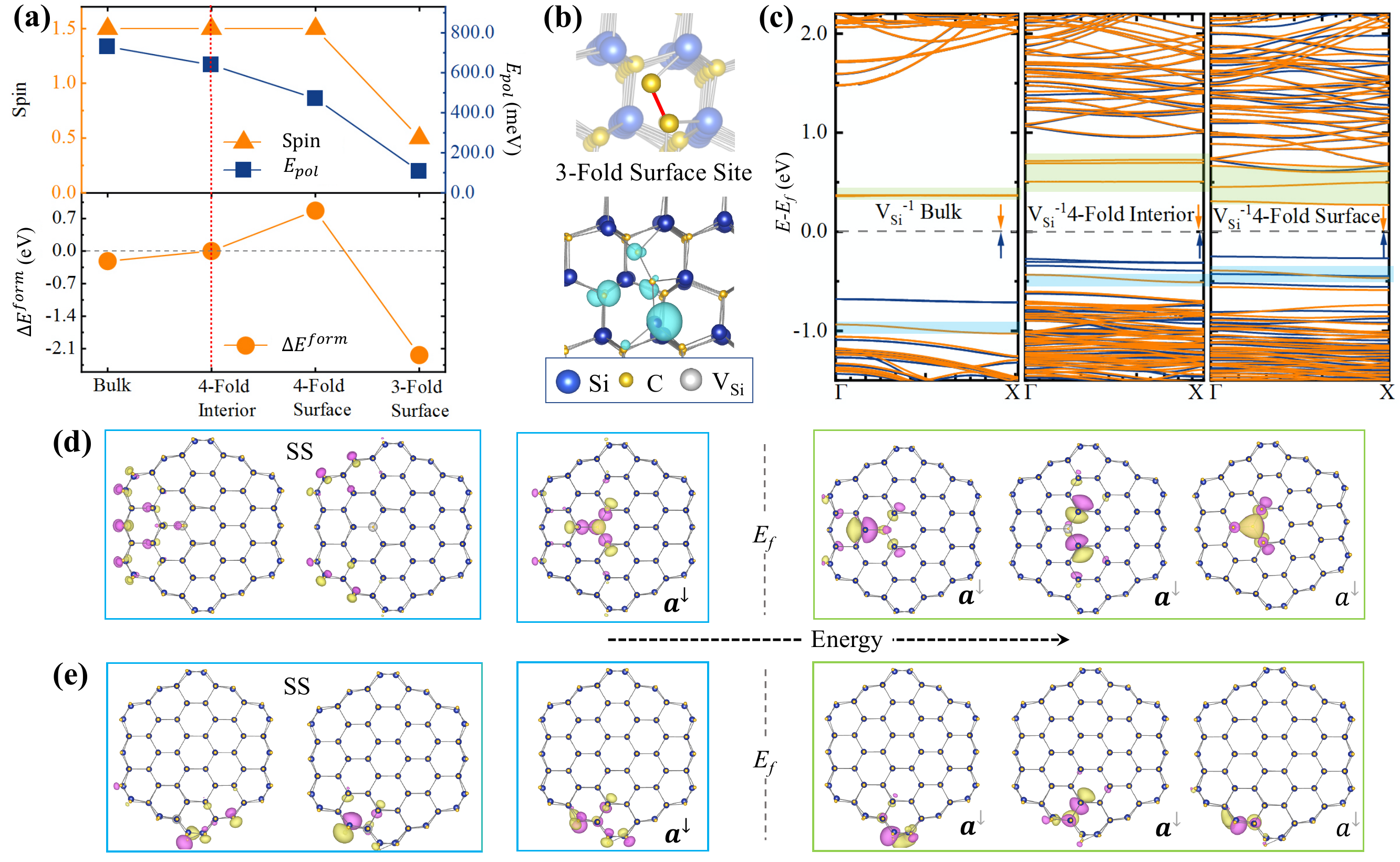}
	\caption{Negatively-charged silicon vacancy, $\text{V}_{\text{Si}}^{-1}$: (a) Site-dependent spin, spin polarization energy ($E_{Pol} = E_{Non-Magnetic} - E_{Magnetic}$), and formation energy, $\Delta E^{form}$, of the defects at different sites in the NW. Defect formation energy at the interior of the wire is used as reference. (b) Equilibrium geometry for $\text{V}_{\text{Si}}^{-1}$ at the 3-fold coordinated surface site, showing dimerization of carbon atoms. The lower figure is a plot of spin density difference (in blue), showing the distribution of the defect's spin on atoms around $\text{V}_{\text{Si}}^{-1}$. (c) Band structures (from left to right) for $\text{V}_{\text{Si}}^{-1}$ created in bulk 2H-SiC, at the NW center, and the four-fold coordinated site at the NW surface, respectively. The optically-active (spin-down channel) defect states are highlighted. The Fermi energy, $ E_{f}$, is used as the reference. Charge density plots for the optically-active defect states (energy-ordered) for the defect at: (d) the NW center, and (e) the 4-fold coordinated site on the NW surface. Also shown are donor-like surface states (SS) that are just below the filled defect states and show hybrid defect-surface states character. Pink (yellow) color corresponds to positive (negative) isovalues.}
	\label{fig:VSiqm1}
\end{figure*}

\subsection{Negatively charged silicon vacancy in the NW}  

The equilibrium structure of the pristine NW [Fig.~\ref{fig:SiC-NW-Structure}(a)] shows the characteristic contraction of bonds between atoms on the surface of nanowire~\cite{Huang2008SkinEffect}. This contraction is mostly driven by the surface effects since the chemical environment of the atoms on the surface is different from that of the underlying bulk. This results in a coordination-dependent reconstruction of bonds on the NW surface. For the under-coordinated surface sites, the three remaining bonds are strengthened, with the average basal plane bonds contracting by about 3.20\%, and the bond along the axial direction showing a contraction of about 8.50\%. The 4-fold coordinated surface-sites show a much smaller contraction, where the bond lengths change to accommodate structural changes imposed by the neighboring 3-fold coordinated sites, with which they share atoms. For these sites, the bond lengths for atoms on the basal plane contract by about 0.65\%, and the bond along the axial direction contracts by about 0.50\%. Moving inwards towards the NW center, different sites show changes in bond lengths to accommodate structural changes at the surface. For example, the 4-fold interior site used in the present study shows a slight increase in bond lengths -- about 0.15\% and 0.76\% increases in the axial- and basal-directions, respectively. The tendency of the nanostructures to reduce surface energy via skin-bond contraction is expected to affect different defect properties. 

The site-dependent spins, spin polarization energies and defect formation energies of $\text{V}_\text{Si}^ {-1}$ in the SiC NW are plotted in Fig.~\ref{fig:VSiqm1}(a). Properties of defects in the bulk are provided for comparison, where appropriate. 
The net spin remains 3/2 at a site in the NW interior and the four-fold coordinated site on the surface of the NW, which is consistent with the bulk calculations. On the other hand, ${\text{V}_\text{Si}}^ {-1}$ at the under-coordinated site is a spin-1/2 defect. Site-dependence is also observed in the spin polarization energy, which is defined as the energy difference between the magnetic and non-magnetic structures ($E_{Pol} = E_{Non-Magnetic} - E_{Magnetic}$).  For all sites, the calculated values of $E_{Pol}$ are substantially larger than the room temperature thermal energy ($\mathrm{k_{B}T\approx 25\,meV}$), showing that the spins in not only the bulk, but also in the NW will survive well beyond room temperature. The defect formation energies at different sites in Fig.~\ref{fig:VSiqm1}(a) are given using the formation energy in the center of the NW as the reference.  As expected, the formation energies in the bulk and the interior of the NW are comparable. On the other hand, the formation energy at the 4-fold coordinated surface site on the surface is higher than that in the interior of the NW, even though the creation of the silicon vacancy requires breaking four covalent bonds for both sites. This can be traced back to the contraction of the bonds on the surface of the NW, with a concomitant increase in bond strengths on the surface.  For the 3-fold coordinated site, we find that the formation energy is substantially smaller (by $\sim$2.23\,eV) as compared to the defect in the center of the NW [see Fig.~\ref{fig:VSiqm1}(a)].  The decrease in the defect formation energy at the undercoordinated site can be attributed to: (i) having to break only three strong covalent bonds at a three-fold site as compared to any other site where the silicon is tetrahedrally bonded to carbon-atoms and (ii) large reconstruction at the undercoordinated site, once $\text{V}_\text{Si}^ {-1}$ is created. Figure~\ref{fig:VSiqm1}(b) shows this reconstruction at the defect site, with two of the three carbons around the defect dimerizing and forming a chemical bond with each other, thereby reducing the energy of the structure considerably. The bonding of the carbon atoms at the defect site also reduces the net spin of $\text{V}_\text{Si}^ {-1}$ to a value of 1/2 instead of the expected spin-3/2 value obtained for the defects in 4-fold coordinated sites in the bulk and NW. The spin density difference plotted in Fig.~\ref{fig:VSiqm1}(b) shows that the 1/2-spin of this defect is distributed on different atoms around the defect, with large contributions coming from nearby Si- and C-atoms.  Hence, $\text{V}_\text{Si}^ {-1}$ at the 3-fold site loses its desirable properties, and is no longer usable as a spin-3/2 qubit.

Figure~\ref{fig:VSiqm1}(c) shows the bandstructures (from left to right)  for the $\text{V}_{\text{Si}}^{-1}$-defect created in the bulk 2H-SiC, at the NW center, and the four-fold coordinated site at the NW surface. The nearly-dispersionless defect states belonging to the optically-active spin-down channel are highlighted. Unlike bulk 2H-SiC, where the $\tilde{a}_{1}$- and $e$-states are almost degenerate, there is a complete loss of degeneracy of the unoccupied defect states in the NW as can be seen in Fig.~\ref{fig:VSiqm1}(c).  This is due to the much lower symmetry ($C_{1}$) of the $\text{V}_{\text{Si}}^{-1}$-defects in the NW. Thus, for $\text{V}_{\text{Si}}^{-1}$ at the 4-fold coordinated sites, there can be several possible spin-preserving and dipole-allowed optical excitations between the defect states. In the case of the 4-fold coordinated interior-site, we find that there are three possible excitations from the filled spin-down singlet state to the three empty defect states. The charge density isosurfaces of each of these states involved in the excitations are shown in Fig.~\ref{fig:VSiqm1}(d), where pink- and yellow-colors correspond to the positive and negative isovalues, respectively. The lowest-energy photoexcitation involves the highest filled spin-down defect state (belonging to ($a$-representation) to the lowest empty spin-down singlet state, which also has $a$-symmetry. The corresponding ZPL is calculated to be 0.87\,eV, which is much lower than the ZPL in the bulk. In addition to the in-gap filled defect state, there are two nearly-degenerate filled states (spin-down channel) that lie about 0.17\,eV below the highest filled defect state. These donor states are mostly-localized on the surface, though one of them shows some defect-state character. The surface states (SS) may participate in the photoexcitation process. Supplemental Fig. 1 shows these in-gap states (surface- and defect-states) for a smaller isovalue, further highlighting their nature and the extent of hybridization between the defect states and the surface-states.

In the case of the 4-fold coordinated surface-site, we find three filled spin-down (optically-active spin channel) in-gap states, which are spatially localized. They show a hybrid defect-surface character, with carbon atoms surrounding the defect, as well as the adjacent surface atoms, contributing to these states [see Fig.~\ref{fig:VSiqm1}(e)]. The two lower in-gap states lie below the highest filled state by about 0.44\,eV  and 0.16\,eV. All three of the in-gap states may participate in the photoexcitation process. The ZPL energy for the lowest-energy optical excitation, involving the highest filled spin-down singlet defect state to the lowest empty spin-down singlet state, is calculated to be 0.70\,eV, which is smaller than the ZPL-values calculated in the bulk and interior of the NW.  We also find that highest empty spin-down defect state is resonant with the surface states close to the conduction band edge. Electrons excited to these surface states via off-resonant excitations or via two photon processes may lead to either blinking, if the surface states act as the shelving state~\cite{Kaviani2014}, and/or photobleaching of the defect with a permanent loss of electron upon excitation.  Hence, we find that upon photoexcitation, there are a number of processes that can result in a change from the bright to dark charge state~\cite{Wolfowicz_Dark_State_PNAS2018} of the defect, converting $\text{V}_{\text{Si}}^{-1}$ to $\text{V}_{\text{Si}}^{0}$.  
As the latter has unknown photoluminescence properties and its electronic structure properties have  remained somewhat controversial~\cite{Zywietz_JT_PRB_1999,VSiq0_Spin1_Torpo1999,Vsiq0_Gali_1999}, we present its electronic and spin properties in Supplemental Note I and Supplemental Fig. 2. Our results show that the defect in bulk SiC prefers to be in a high spin state ($\text{S}=1$), in agreement with the Zywietz \textit{et al.}~\cite{Zywietz_JT_PRB_1999} and Torpo \textit{et al.}~\cite{VSiq0_Spin1_Torpo1999}. 

\begin{figure*}
 	\centering
 	\includegraphics[width=0.75\linewidth]{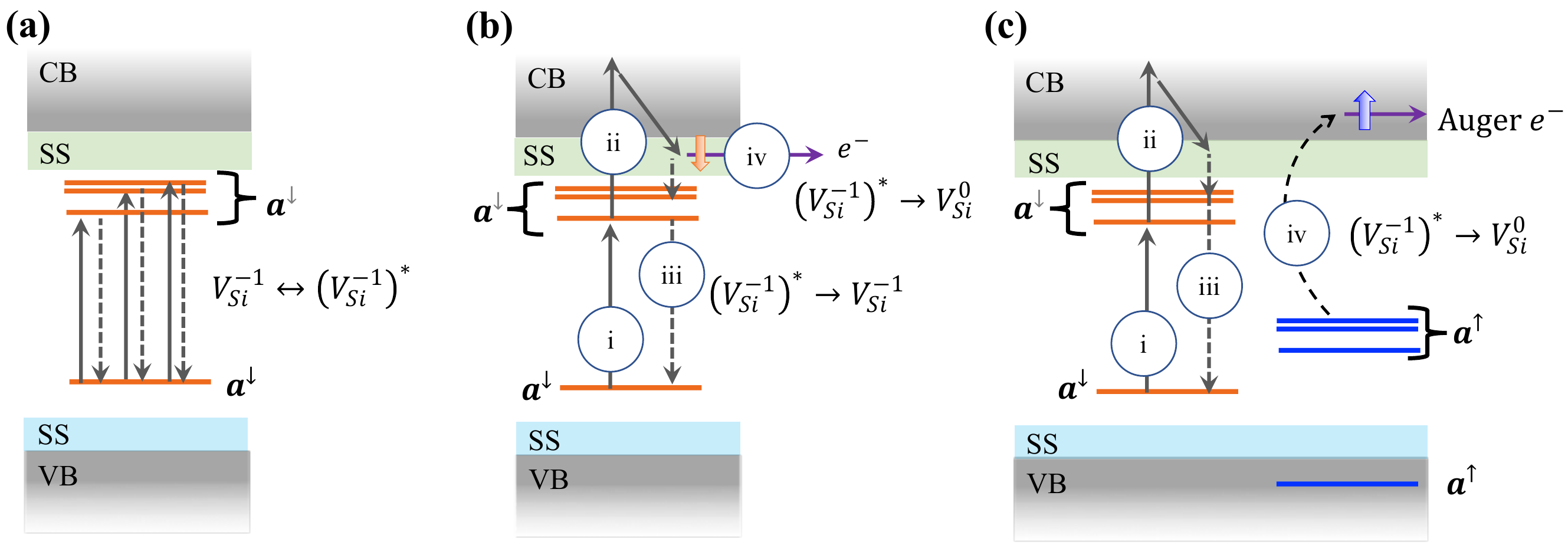}
 	\caption{Optical excitation and charge conversion of $\text{V}_{\text{Si}}^{-1}$. (a) Schematic diagram (not to scale) showing photo-excitation [$\text{V}_{\text{Si}}^{-1} \rightarrow (\text{V}_{\text{Si}}^{-1})^{\ast}$] and de-excitation [$(\text{V}_{\text{Si}}^{-1})^\ast \rightarrow \text{V}_{\text{Si}}^{-1}$] resulting in photoluminescence from the $\text{V}_{\text{Si}}^{-1}$-defect in the NW. Only the filled and empty minority-spin defect states (all belonging to $a$-representation) are shown. $(\text{V}_{\text{Si}}^{-1})^{\ast}$ refers to the excited state of the defect. The donor- and acceptor-type surface states (SS) are indicated. (b) Schematic diagram showing excitation of an excited defect [i.e. \circled{i} followed by \circled{ii}] in a two-photon absorption process. Scattered into the localized acceptor-like surface state, the electron can either undergo phosphorescence [marked as  \circled{iii}], resulting in blinking, or it can migrate away from the defect, resulting in a charge conversion from $\text{V}_{\text{Si}}^{-1} \rightarrow  \text{V}_{\text{Si}}^{0}$, with the latter being a dark state. (c) Schematic diagram showing another possible route to charge conversion to the neutral state, involving a two-photon absorption process [\circled{i} followed by \circled{ii}]  followed by an Auger process [marked by \circled{iii} and \circled{iv}] that ionizes the defect. Here, the filled majority spin-states are also shown because spin-up electron is ionized from the defect. }
 	\label{fig:VSiqm1_chargeConversion}
 \end{figure*}

\section{Discussion}

The potential loss of signal from the $\text{V}_{\text{Si}}^{-1}$-defects close to the surfaces of the nanostructured host will be deleterious to the performance of the devices based on these defects and requires further understanding so that mitigating protocols can be designed. Our calculated bandstructures for the monovacancies in the NW [Fig.~\ref{fig:VSiqm1}(c)] show the presence of donor and acceptor-type surface states, which are close to the defect states in energies. In addition, due to the proximity of the $\text{V}_{\text{Si}}^{-1}$-defects considered here to the surfaces, the defect states show some hybridization with the surface states. As a consequence, these surface states may become involved in the photoexcitation processes in several ways, which may lead to temporary or permanent loss of the charge from the defect. Examples of different photoexcitation processes that may lead to luminescence and/or charge conversion to a dark state are illustrated in Figs~\ref{fig:VSiqm1_chargeConversion}(a)-(c). Figure~\ref{fig:VSiqm1_chargeConversion}(a) is a schematic diagram showing the processes of photo-excitation [$\text{V}_{\text{Si}}^{-1} \rightarrow (\text{V}_{\text{Si}}^{-1})^{\ast}$] and de-excitation [$(\text{V}_{\text{Si}}^{-1})^\ast \rightarrow \text{V}_{\text{Si}}^{-1}$], resulting in photoluminescence from the $\text{V}_{\text{Si}}^{-1}$-defect in the NW. Here, $(\text{V}_{\text{Si}}^{-1})^\ast$  refers to a photoexcited defect. The levels are not drawn to scale and thus apply to all 4-fold coordinated sites in the NW considered here. Also, both the donor-type and acceptor-type surface states localized on the surfaces are shown as thin colored bands. Figure~\ref{fig:VSiqm1_chargeConversion}(b) shows a two-photon absorption process [\circled{i} followed by \circled{ii}].  The electron can then be temporarily trapped by the surface states, which may act as shelving states~\cite{Kaviani2014}, and then undergo phosphorescence [marked as  \circled{iii}], resulting in blinking. Alternatively,  the electron can migrate away from the defect, resulting in a charge conversion from $\text{V}_{\text{Si}}^{-1} \rightarrow  \text{V}_{\text{Si}}^{0}$, with the latter being a dark charge state.  Figure~\ref{fig:VSiqm1_chargeConversion}(c) shows another pathway for charge conversion to a neutral state, involving a two-photon absorption process [steps \circled{i} and \circled{ii}]  followed by an Auger process in which energy released in step \circled{iii} is used to ionize an electron away from the defect in step \circled{iv}~\cite{Golter_SwitchChargeState_2017}.  Hence, the photoexcitation process itself can result in signal deterioration, with a temporary or permanent loss of an electron from the photo-excited defect. Our calculations also reveal  a second potential source for the loss of signal from a defect close to the NW surface. As discussed earlier, the defect formation energies at the undercoordinated surface sites are much lower than at any other site in the NW. Hence, the defect itself can migrate to the thermodynamically-favorable undercoordinated surface sites, where the defects have very different structural, electronic and spin properties.

We would also like to note that $\text{V}_{\text{Si}}^{-1}$ has a stable spin state at different 4-fold coordinated sites within the NW, irrespective of the local strain at the sites.  There is, however, a site-dependent shift in the ZPL for the lowest energy excitation of the $\text{V}_{\text{Si}}^{-1}$ in the NW as compared to the bulk.  The calculated values of ZPL for the 4-fold sites in the interior and on the surface of the NW are about 325.6\,meV and 491.1\,meV lower than the bulk value, respectively.  The lowering of the ZPL value at the four-fold sites in the NW can be attributed to several mutually-dependent effects, including: (i) local strain due to the skin-contraction effect, (ii) large structural changes in the NW upon photoexcitation, (iii) the hybridization between the surface states and the defect states, and (iv) lowering of symmetry compared to bulk, which splits the unoccupied defect states, reducing the gap between the frontier orbitals (i.e. the highest-occupied and lowest-unoccupied defect states) involved in the lowest energy photoexcitation.  The relative importance of these four factors can be gauged from the DFT results.
Estimating the strains by the changes in the bond lengths at the respective sites, and using the experimentally deduced value of strain-coupling parameter of 1130\,meV(strain$)^{-1}$ (in the basal direction)~\cite{nanostr_VSi_expt_2020}, we can obtain an order of magnitude estimate of the ZPL shift due to strain. We find that at all 4-fold sites in the NW, the strain-induced changes in the ZPL (as compared to the bulk) are about 10\,meV or less.  Although our calculated strain-dependent changes in the ZPL agree with the experimentally observed shifts~\cite{nanostr_VSi_expt_2020}, they are two orders of magnitude smaller than the total ZPL changes calculated for the 4-fold coordinated sites within the NW. This means that the local strain is not the leading cause of the changes in the ZPL  for different sites in our NW.  This conclusion is different from that reached by V\'{a}squez \textit{et al.}~\cite{nanostr_VSi_expt_2020}, where the authors surmised that the strain is the leading cause of the ZPL shift in their experimental study of strain-effects within SiC nanostructures.  We would like to emphasize that these different conclusions do not necessarily mean that there is a contradiction. It is most likely that the difference in conclusions reached in our theoretical work and the experiment is due to the large sizes (few hundreds of nm to a few $\mu$m) of their nanoparticles, with defects that were located much farther away from the surfaces of their nanoparticles than in our study. In turn, this would have minimized the importance of other surface effects in experiment that play a prominent role in our study.  In order to quantify the effect of the second factor -- structural changes upon photoexcitation -- we calculated the Stokes shift, which is given by the difference in the vertical excitation and the ZPL. We find that the Stokes shift is about 104.9\,meV, 138.6\,meV and 169.7\,meV in the bulk, at the site in the NW interior and the 4-fold site on the NW surface, respectively. In turn, it means that the monovacancies in the NW undergo larger structural changes upon photoexcitation.  However, the ZPL shift attributable to the changes in Stokes shift in the NW as compared to the bulk is still an order of magnitude smaller.  The third factor that causes ZPL-shift is the hybridization between defect states and the surface states of the NW.  This is especially pronounced in the defect at the surface. On one hand, there is a hybridization between the donor-like localized surface states and the filled defect states, forming bonding and antibonding MOs. The antibonding MOs, that are mostly defect-state in character are pushed up within the bandgap. On the other hand, the empty defect states that hybridize with the acceptor-like surface states, are pushed down in energy. This reduction in the gap between filled and empty defect states would then lead to the ZPL shift. Lastly, the lowering of symmetry itself at different 4-fold coordinated sites in the NW as compared to the bulk split what were nearly-degenerate empty defect states. This further reduces the gap between the frontier orbitals. The influence of the last two factors can be seen in the energy splitting between the frontier orbitals that we calculate to be about 1287.7\,meV, 938.3\,meV and 833.3\,meV for the $\text{V}_{\text{Si}}^{-1}$ in bulk, at the NW interior site and the NW surface site, respectively. Hence, we find that the energy gap reduction in the frontier orbitals of $\text{V}_{\text{Si}}^{-1}$ in the NW interior and the NW surface (as compared to the bulk) accounts for the majority of the ZPL shift.


In conclusion, our analysis shows that the spin state of $\text{V}_{\text{Si}}^{-1}$ remains stable at different 4-fold coordinated defect sites in the NW, irrespective of the local strains. However, the optical properties of the quantum emitters show profound changes due to finite-size effects.  Further, we find that the electronic, spin and optical properties of the 3-fold sites are completely different, with the defects at these sites adopting a low-spin state. Also, due to considerable reconstruction around the undercoordinated defect-sites, and the fact that fewer covalent bonds are needed to be broken in creating the silicon vacancies, these sites have much lower defect formation energies. These findings suggest that thermodynamically, surfaces might act as a sink for the defects, with the defects at these sites no longer operating as bright defects.  We also show that the photoexcitation process itself might result in deterioration of the signal from the defect due to the participation of the surface states. Such defects will be unusable for different applications designed around their spin-state, and our ability to control and manipulate these spins will be lost over time. This work is an important step towards understanding experiments and devising mitigating protocols which may include surface passivation and better positioning of the implanted defects. In addition, with a better understanding of the finite size effects, one can use nanostructuring as a tool to tune optical properties of the defects themselves.

\vspace{12pt}
\begin{acknowledgments}

\noindent \textbf{Acknowledgements:} We acknowledge support by the National Science Foundation under NSF grant number DMR-1738076 and the STC Center for Integrated Quantum Materials under NSF Grant number DMR-1231319. The computational support was provided by the Extreme Science and Engineering Discovery Environment (XSEDE) under Project PHY180014, which is supported by National Science Foundation grant number ACI-1548562.  For three dimensional visualization of crystals and volumetric data, use of VESTA 3 software is acknowledged. 


\end{acknowledgments}

\noindent \textbf{Author Contributions:} P.D. conceived and directed the study. T.J. performed the ground state calculations.  P.D. performed the CDFT calculations and wrote the final manuscript, with inputs from the initial draft from T.J.  Both authors proof-read and reviewed the final revision of the manuscript.

\normalem  
\bibliography{biblio}
\bibliographystyle{apsrev}

\end{document}